\documentclass[5p,sort&compress,times,letterpaper]{elsarticle}

\usepackage{amsmath,amssymb}
\usepackage{color}
\usepackage[utf8]{inputenc}
\usepackage{graphics}      
\usepackage{graphicx}      
\usepackage{longtable}     
\usepackage{url}           
\usepackage{bm}            
\usepackage{xcolor}
\usepackage{float}
\usepackage[colorlinks,allcolors=blue]{hyperref}

\begin{document}

\title{On Fast Neutrino Flavor Conversion Modes in the Nonlinear Regime}

\author{Sajad Abbar}
\author{Maria Cristina Volpe}
\address{Astro-Particule et Cosmologie (APC), CNRS UMR 7164, Universite Denis Diderot, 10, rue Alice Domon et Leonie Duquet, 75205 Paris Cedex 13, France }

\begin{abstract}
A dense neutrino medium could experience self-induced flavor conversions on relatively
small scales in the presence of the so-called \textit{fast flavor conversion modes}. 
Owing to the fact that fast  conversion scales  could be much smaller
than the ones of the traditional collective neutrino oscillations,
it has been  speculated that fast modes could lead to some sort of 
flavor decoherence/equilibrium.
We study the evolution of fast modes in the nonlinear regime and we 
show that
not only fast modes are not guaranteed to lead to flavor equilibrium,
but also they could lead to some sort of  collective neutrino oscillations
but on scales determined by neutrino number density. 
For the $\nu_e$ dominated case, 
we observe large amplitude 
collective oscillations, whereas a sort of flavor stabilization is reached for 
the $\bar{\nu}_e$ dominated case.
\end{abstract}

\maketitle

 \section{Introduction} 
The most extreme astrophysical 
sites such as core-collapse supernovae and 
 neutron star mergers  include  dense neutrino media.
 The nature of the neutrino evolution in such a dense neutrino
 gas could be very different from the one in vacuum and matter.
 Indeed, due to the presence of neutrino-neutrino interactions,
 the evolution of neutrinos in dense neutrino media is 
 a nonlinear correlated problem \cite{Pantaleone:1992eq, sigl:1992fn, 
 duan:2006an}.   

The first studies on this problem were carried out in  simplified 
symmetric models such as the stationary spherically symmetric
 \textit{neutrino bulb model} \cite{duan:2006an, duan:2006jv,
 duan:2007sh}.
  The most prominent feature observed in these studies 
was the presence of collective neutrino oscillations where neutrinos
 evolve collectively due to the correlation induced by neutrino-neutrino
interaction. 
The most remarkable observational consequence of this collective
 evolution is the well-known spectral swapping in which $\nu_e$ ($\bar{\nu}_e$)
 exchange its spectrum with $\nu_x$ ($\bar{\nu}_x$) for a range of neutrino 
 energies \cite{duan:2006jv, duan:2007sh, dasgupta:2009mg, duan:2010bg,
 Galais:2011gh, Duan:2007bt}.

It was just more recently that it was realised the spatial/time
 symmetries 
 could be broken spontaneously in a dense neutrino gas
 \cite{raffelt:2013rqa, duan:2013kba, duan:2014gfa,
 abbar:2015mca, Abbar:2015fwa, chakraborty:2015tfa, Chakraborty:2016yeg,
 Dasgupta:2015iia, Mirizzi:2015fva}. 
This, in principle, could lead to new neutrino flavor conversion phenomena.
On the one hand, the breaking of the spatial symmetry allows for 
neutrino flavor conversion at very large neutrino number densities.
On the other hand, the breaking of time symmetry could remove 
the matter suppression of neutrino oscillations and might lead to  
flavor conversion at very large matter densities.

Another important development was the discovery of fast flavor conversion
modes that could occur on very small scales 
\cite{Sawyer:2005jk, Sawyer:2008zs, Sawyer:2015dsa,
 Chakraborty:2016lct, Izaguirre:2016gsx, Wu:2017qpc, Tamborra:2017ubu,
  Capozzi:2017gqd,
 Dasgupta:2016dbv, Abbar:2017pkh, Dasgupta:2018ulw, Airen:2018nvp}. 
Unlike the case
of the traditional collective (slow) modes which do occur on scales determined
by  neutrino vacuum frequency $\omega = \Delta m_{\mathrm{atm}}^2/2E $ (which is
 $\sim  \mathcal O  (1)$ km for a $10$ MeV neutrino),
 fast modes  occur 
 on scales $\sim G_{\mathrm{F}}^{-1} n_\nu ^{-1}$
 with $n_\nu$ being the neutrino number density.
 Obviously, the conversion scales for fast modes could
 be much smaller than the ones for slow modes at large enough
 neutrino number densities.
 It has also been argued that the presence of crossing in the angular distribution of
 electron lepton number carried by neutrinos ($\nu$ELN)  is
 a necessary condition for the occurrence of  fast modes
 \cite{Izaguirre:2016gsx, Capozzi:2017gqd,
 Dasgupta:2016dbv, Abbar:2017pkh}. 
 In particular, it has been shown that the presence of crossing(s) in $\nu$ELN 
 could allow $G_{\rm{F}} n_\nu$ to play the role of $\omega$
 \cite{Abbar:2017pkh}.
 
Not only is fast modes
 an amazing phenomenon by itself,
but also it could have important implications for the physics
of supernovae. Firstly, it could help
removing matter suppression by activating
unstable neutrino modes (with large frequencies in time) on small
enough scales. Secondly, it could cause neutrino flavor conversion
within SN regions that have long been thought to be the realm
of scattering processes. In fact, this is expected
since  fast modes do occur on scales  
$\sim G_{\rm{F}}^{-1} n_{\nu} ^{-1}$,
whereas  scattering processes occur on scales
 $\sim G_{\rm{F}}^{-2} E^{-2} n_{\rm{B}} ^{-1}$
 with $n_{\rm{B}}$ being the baryon number density 
 \cite{Cirigliano:2017hmk, Capozzi:2018clo}.
 Within the neutrino decoupling region, 
the former could have values $\lesssim \mathcal O  (1)  $ cm, 
whereas the latter could be much larger 
 $\gtrsim \mathcal O  (1) $ km.

Since   fast conversion modes could occur on scales much smaller
than the ones of the traditional collective modes,
it has been widely speculated that fast modes could lead to some sort of 
flavor decoherence/equilibrium. Needless to say, the occurrence of  flavor equilibrium
could significantly simplify the physics of supernova neutrinos 
from both theoretical and observational points of view. On the theoretical side,
it could remarkably reduce the computational difficulties. On the observational side,
it could notably improve the analysis of the supernova neutrino signals.

Although the total flavor
equilibrium might be inaccessible due to the lepton number conservation
laws, some sort of partial flavor equilibrium (decoherence) has been proposed
to occur.
This could be very similar to the evolution of decoherence
for small $\nu_e$-$\bar{\nu}_e$ asymmetries
 ($\alpha = n_{\bar{\nu}_e}/n_{\nu_e}$ close to one)
 within 
slow modes,
 where $\bar{\nu}_e$ 
could experience flavor equilibrium 
and full decoherence\footnote{This is the case for $\alpha<1$. For
the $\bar{\nu}_e$ dominated case, it might be  $\nu_e$ that experiences equilibrium.} 
 with survival probability $P_{\bar{\nu}_e \bar{\nu}_e} \approx 1/2$ 
\cite{Raffelt:2007yz, EstebanPretel:2007ec, Raffelt:2010za}. 
Likewise, $\nu_e$ 
could experience decoherence up to the extents allowed by the conservation
laws.

In this work, we study fast neutrino flavor conversion modes in the nonlinear
regime by using a one dimensional spatial/temporal schematic model. We show that not only fast
modes are not guaranteed to lead to flavor equilibrium,
but also they could lead to large amplitude collective neutrino oscillations 
on very small scales for $\alpha<1$. Moreover, for $\alpha>1$, 
an out of flavor equilibrium stabilization could be reached.

\section{THE NEUTRINO LINE MODEL}  
\begin{figure} [t]
\begin{center}
\includegraphics*[width=.42\textwidth,trim=20 70 20 80,clip]{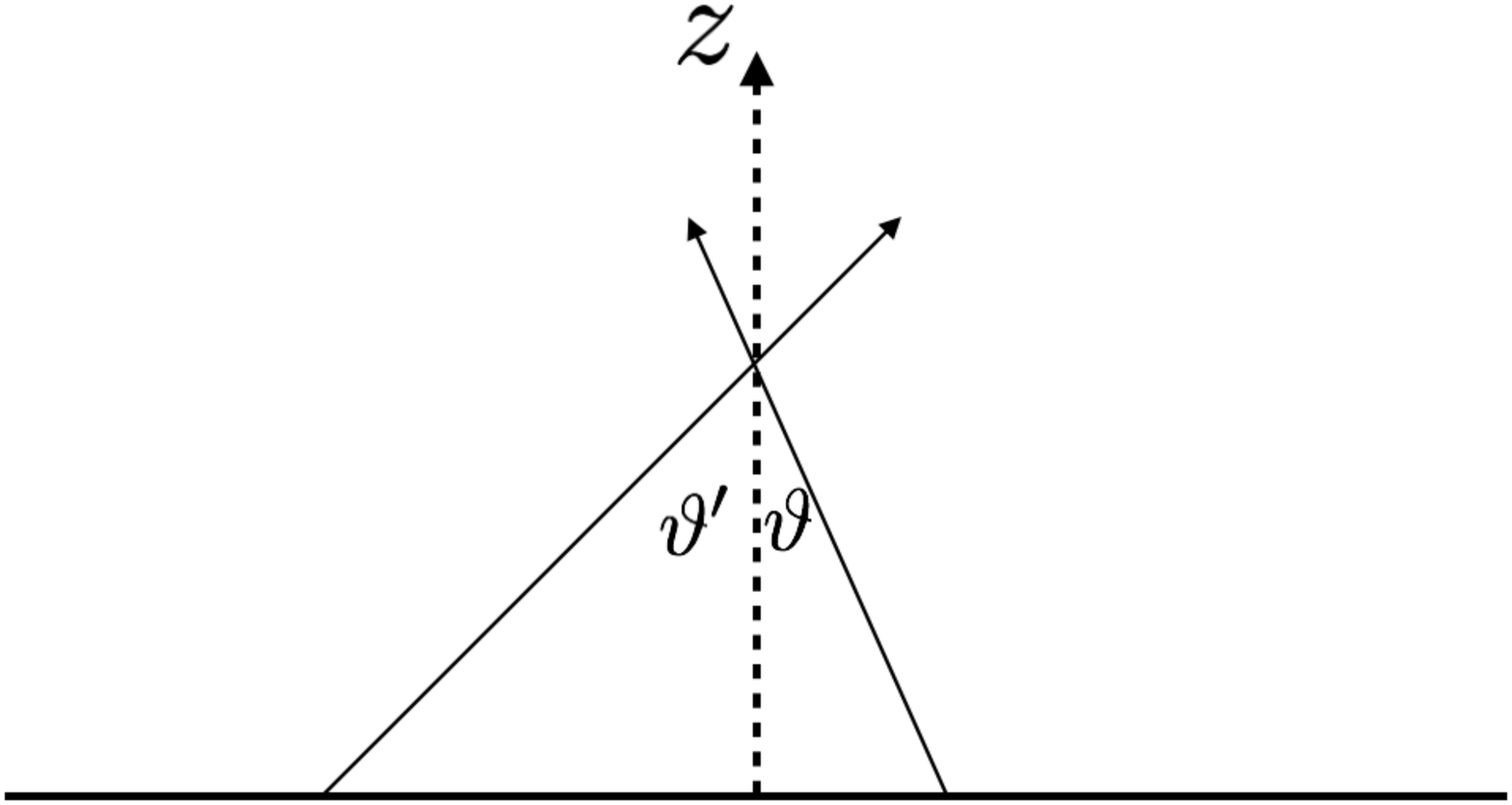}
\end{center}
\caption{A schematic representation of the neutrino
Line model. Neutrinos are emitted from the neutrino line  
with emission angles $\vartheta$ within the range $[-\vartheta_{\rm{max}}, \vartheta_{\rm{max}}]$.} 
\label{fig:geom}
\end{figure}

To study the spatial evolution of fast modes in the simplest multiangle configuration 
(the temporal evolution will be discussed later in this paper),
we consider a stationary one dimensional schematic model
in a two-flavor scenario in which 
\textit{electron} neutrinos and antineutrinos are emitted from 
an infinite line with
emission angles in the range 
$[-\vartheta_{\rm{max}}, \vartheta_{\rm{max}}]$ (Fig. \ref{fig:geom})
\cite{duan:2014gfa, abbar:2015mca}.
We also assume that the symmetry is preserved in the transverse
direction (along the line).

It is  assumed that neutrinos and antineutrinos are 
emitted monochromatically \footnote{This assumption is
made for the sake of definiteness. Otherwise, the EOM
should be \emph{approximately} blind to the neutrino frequency in the presence
of fast modes. However, one should note that there still
might exist some subleading effects from $\omega$ \cite{Airen:2018nvp}. }
 and with the normalised 
angular distributions $f_{\nu_e}(\vartheta)$ and 
$f_{\bar{\nu}_e}(\vartheta)$.
In addition, to observe  fats modes 
we allow for different  $f_{\nu_e}(\vartheta)$ and 
$f_{\bar{\nu}_e}(\vartheta)$ so that there can exist 
crossing in the $\nu$ELN.

 At each point $z$, the state of a neutrino 
 which is traveling in direction $\vartheta$ 
  could be specified by its density matrix $\rho_{\vartheta}(z)$,
and, in the absence of collision, its flavor evolution is given by the equation of motion (EOM)
\cite{sigl:1992fn,Volpe:2013jgr, Cirigliano:2014aoa, Hansen:2016klk,
Stirner:2018ojk, Volpe:2015rla, Serreau:2014cfa}
\begin{equation}
i \cos\vartheta \  \partial_z  \rho_{ \vartheta} = [\rm{H}_{\vartheta},\rho_{ \vartheta}],
\label{eq:EOM}
\end{equation}
with $\rm{H}_{\vartheta} = \rm{H_{vac}} + \rm{H_{mat}} + \rm{H}_{\nu \nu, \vartheta}$ 
being the total Hamiltonian where
\begin{align}
\rm{H_{vac}} &= \frac{1}{2}
\left[ {\begin{array}{cc}
-\omega \cos2\theta_{\textrm{v}}  &   \omega \sin2\theta_{\textrm{v}}   \\
 \quad \omega \sin2\theta_{\textrm{v}}   & \omega \cos2\theta_{\textrm{v}}  \\
\end{array} } \right], \\
 \rm{H_{mat}}  &= \frac{\lambda}{2}
\left[ {\begin{array}{cc}
1   &   0  \\
0  & -1 \\
\end{array} } \right], 
\end{align}
are the contributions from vacuum and matter, 
 $\theta_{\textrm{v}}$ is the neutrino vacuum mixing angle,
$\lambda = \sqrt2 G_{\mathrm{F}} n_e$ with $n_e$ being 
the electron number density and 
\begin{equation}
\begin{split}
 \mathrm{H}_{\nu \nu,\vartheta} = \mu \int_{-\vartheta_{max}}^{\vartheta_{max}} 
&[f_{\nu_e}(\vartheta') \rho_{\vartheta'}(z) - \alpha f_{\bar{\nu}_e}(\vartheta') \bar{\rho}_{\vartheta'}(z)]\\
& \times [(1- \cos(\vartheta - \vartheta')]   \mathrm{d}\vartheta' ,
 \end{split}
 \end{equation}
 is the contribution from $\nu-\nu$ interaction where 
  $\mu = \sqrt2 G_{\rm{F}} n_{\nu_e}$.  
 The EOM for antineutrinos is the same except that $\omega \rightarrow -\omega$. 
 Moreover, it is assumed that $\Delta m_{\mathrm{atm}}^2 > 0$ ($ < 0$) in 
 the normal (inverted) mass hierarchy. 
 
 
 Due to the neutron richness of the supernova medium
 and the neutron star merger environment, $\bar{\nu}_e$'s are decoupled 
at smaller radii than $\nu_e$'s. This means that one should naively
expect $\bar{\nu}_e$ to be more peaked in the forward direction
than $\nu_e$.
 In this study,
 we set $f_{\nu_e}(\vartheta)$ to be constant within the range 
 $[-\vartheta_{max}, \vartheta_{max}]$ and 
 \begin{equation}
f_{\bar{\nu}_e}(\vartheta) \propto \exp(- \vartheta^2/ 2\sigma^2)
\end{equation}
with $\sigma^2 = \pi/60$. It should be noted that to observe  fast modes,
 one could also take uniform $f_{\bar{\nu}_e}(\vartheta)$ and $f_{\nu_e}(\vartheta)$  
 with different opening angles for neutrinos and antineutrinos.  
 However, the choice we made here allows for
 a smoother transition from $\bar{\nu}_e$ to $\nu_e$ dominated
 angular range.
Nevertheless, as we will discuss in the next section, the
qualitative features of our results  
do not depend on the choice of angular distributions.

\section{Results and Discussion}

\begin{figure*} [p]
 
 \centering
\includegraphics*[width=.8\textwidth,trim=25 50 10 20,clip]{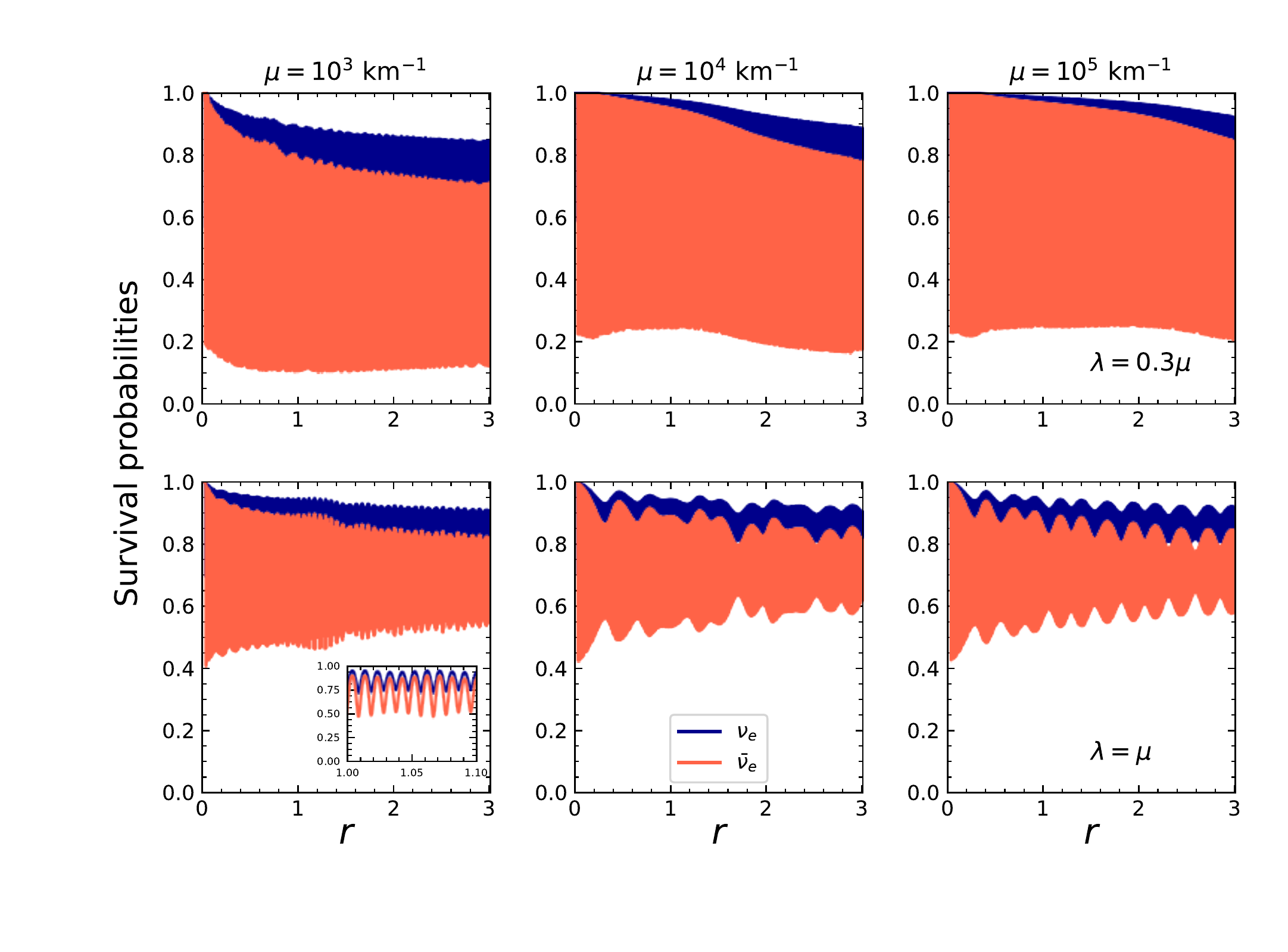}
\caption{ The angle-averaged  survival
probabilities of electron neutrinos (blue line) and electron antineutrinos (red line)
 for $\alpha=0.5$, $\mu= 10^3$, $10^4$ and $10^5$ km$^{-1}$
with $\lambda = 0.3\mu$ (upper panels) and $ \lambda = \mu$ (lower panels)
as a function of the unitless distance defined in the text.
The actual distance is then $z=10^4 r/\mu$ km for each case. 
On the  lower left panel,
we provide a zoomed up subplot of survival probabilities in the range
$r = 1.$ to $1.1$.  
 } 
\label{fig:alpha5}

\includegraphics*[width=.9\textwidth,trim=0 10 10 0,clip]{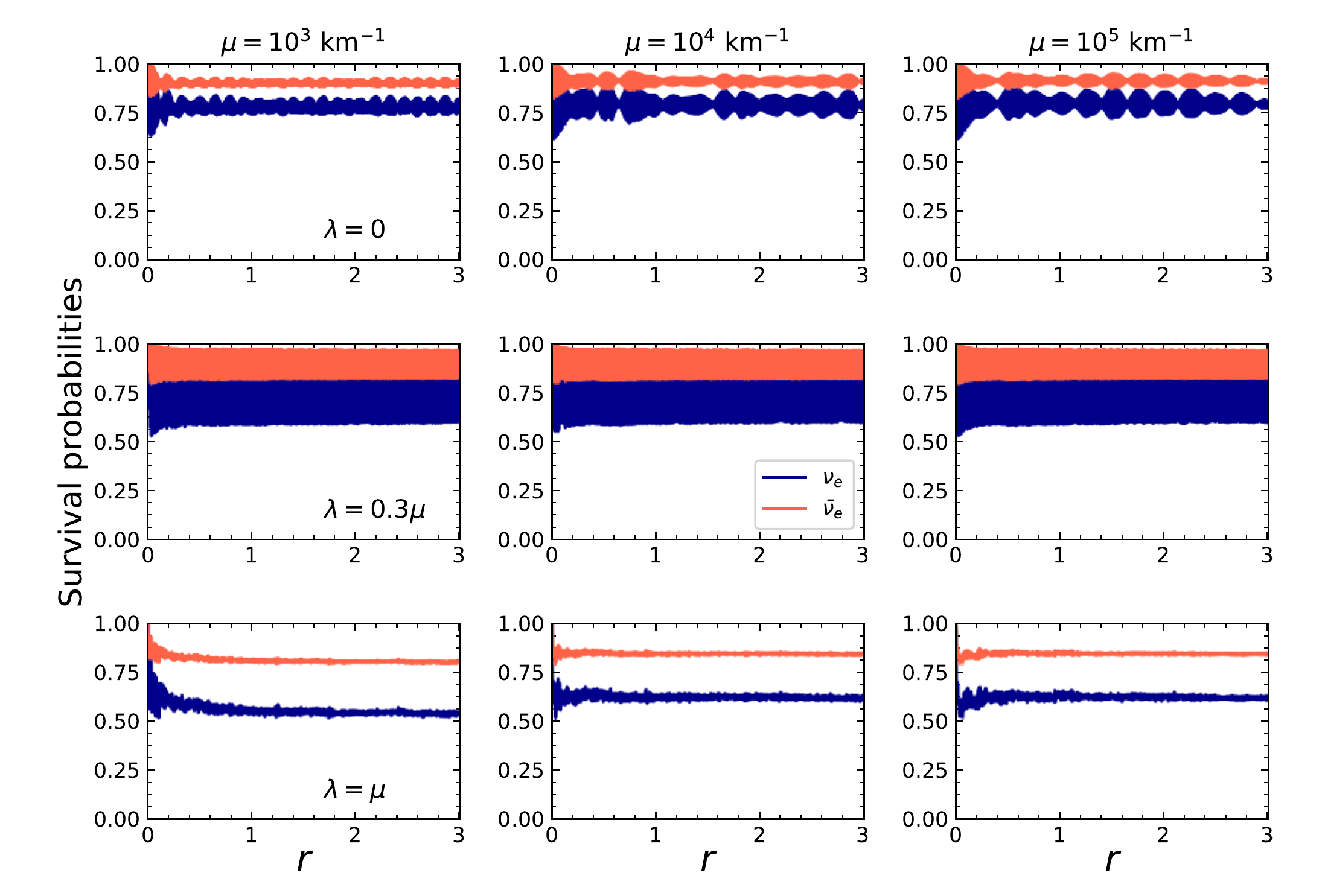}
\caption{The same information as in Fig. \ref{fig:alpha5} for $\alpha=2$ except that three
matter densities $\lambda = 0$, $\lambda = 0.3\mu$ and $\lambda = \mu$
are shown. } 
\label{fig:alpha2}

\end{figure*}

\begin{figure*} [t!]
 \centering
\begin{center}
\includegraphics*[width=.90\textwidth,trim=25 20 10 10,clip]{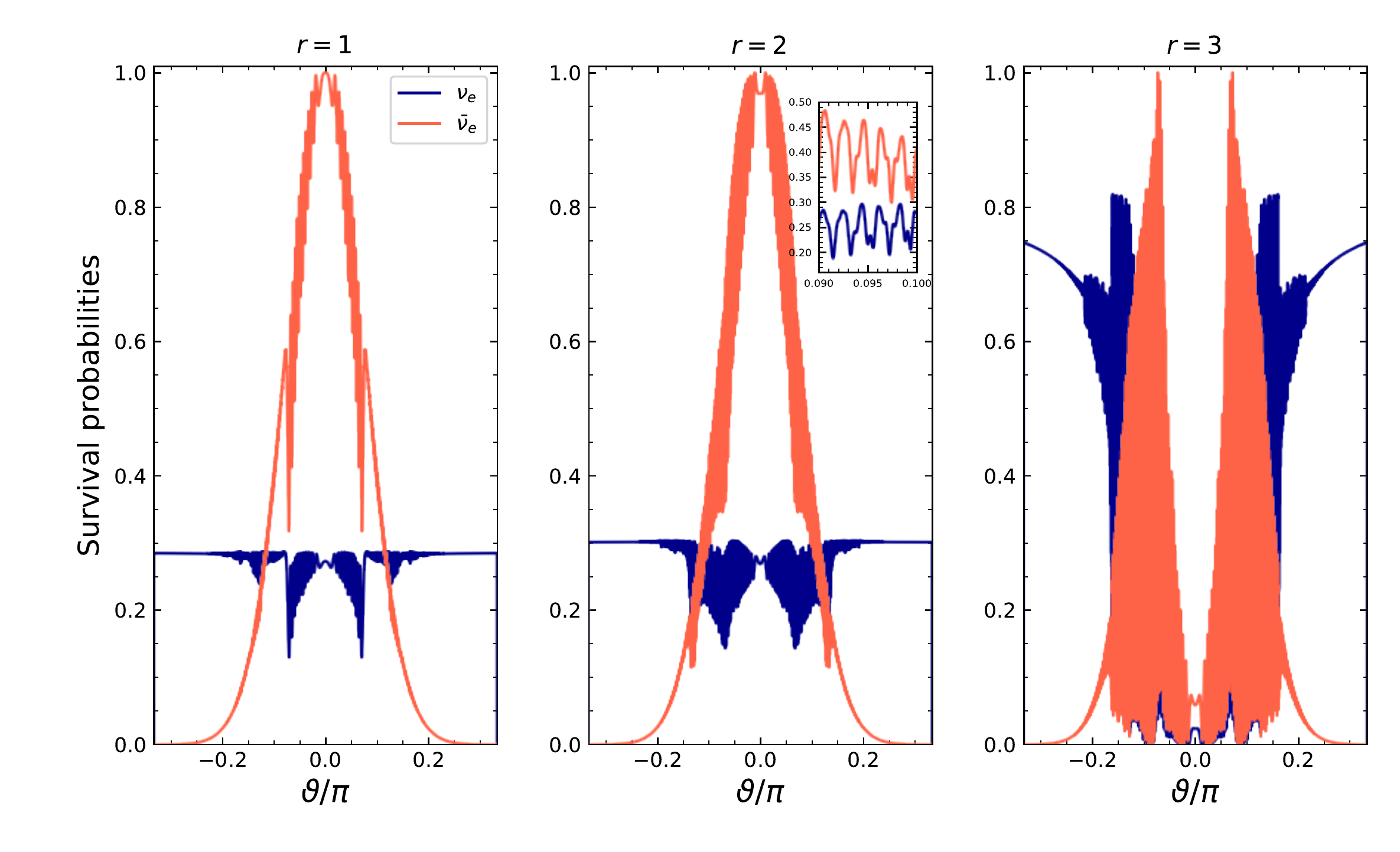}
\end{center}
\caption{  Angular distributions of the survival probabilities of
 neutrinos (blue line) and antineutrinos (red line) 
at different radii.
On the middle panel,
we provide a zoomed up subplot of survival probabilities in the range
$\vartheta/\pi = 0.09$ to $0.1$. } 
\label{fig:angular}
\end{figure*}

\begin{figure*} [t!]
 \centering
\begin{center}
\includegraphics*[width=.95\textwidth,trim=25 05 10 0,clip]{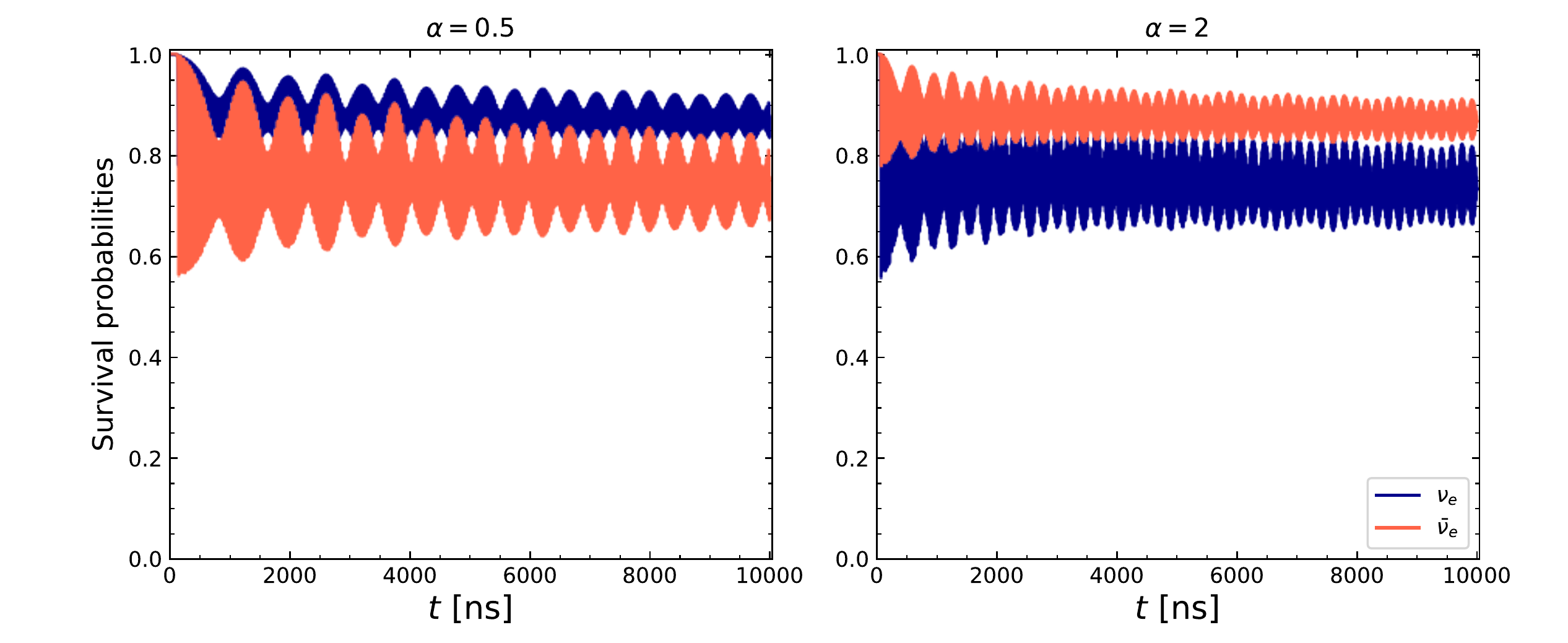}
\end{center}
\caption{  The angle-averaged  survival
probabilities of neutrinos (blue line) and antineutrinos (red line)
 for  $\mu= 10^4$ and $\lambda = 0.3\mu$ with $\alpha=2$ (right panel) and $\alpha=0.5$ (left panel)
as a function of time. Note that as mentoned in the text,  $\alpha=0.5$ calculations
are performed with exchanging initial $\nu_e$ and $\bar{\nu}_e$ angular distributions.} 
\label{fig:time}
\end{figure*}

We have studied fast neutrino flavor conversion modes in the 
nonlinear regime. 
 We assumed $\omega=-1$ (inverted hierarchy), $\vartheta_{\rm{max}} = \pi/3$
 and very small $\theta_{\rm{v}}$ \footnote{
 Note that one can make this choice if the matter density
 is very large. However, one should bear in mind that 
 if the growth rate $\kappa \gtrsim \lambda$,
then it is not physically allowed to assume small effective $\theta_{\rm{v}}$
 since
 one may not find such a rotating frame
  where the vacuum term oscillates "\textit{so}" quickly
  that the off-diagonal term averages   to zero.}. 
 Nevertheless, we have confirmed
  that the results we are presenting here do 
 not seem to depend qualitatively on the choice of
  $\omega$, $\theta_{\rm{v}}$,
  $\vartheta_{\rm{max}}$ 
or even on the details of the shape of $f_{\nu_e}(\vartheta)$ and $f_{\bar{\nu}_e}(\vartheta)$
as long as  fast modes exist. 
We investigated the evolution of fast modes for a number of
$\alpha$'s for both $\nu_e$  ($\alpha<1$) and
 $\bar{\nu}_e$ ($\alpha>1$) dominated cases.
   In Figs. \ref{fig:alpha5} and \ref{fig:alpha2}, 
we show the results for the spatial evolution of neutrinos for $\alpha=0.5$ and $2$
(see Fig. \ref{fig:time} for the temporal case). 
In general, for the $\nu_e$ dominated cases
 we observed large amplitude oscillations, 
   whereas a sort of flavor stabilization could be reached for 
the $\bar{\nu}_e$ dominated case.

In Fig. \ref{fig:alpha5}, we show the angle-averaged neutrino and antineutrino survival
probabilities for $\mu= 10^3$, $10^4$ and $10^5$ km$^{-1}$
and  $\lambda = 0.3\mu$ and $\mu$
as a function of the unitless distance defined by $r=\mu z/10^4$. 
Since the relevant scale for fast modes is $\sim 1/\mu$,
the unitless quantity $r$ provides a physical measure of 
length in the problem.
For these values of $\lambda$, the value of the growth rate is $\kappa \approx 0.05 \mu$. 
The most prominent feature of this plot is the presence of  
large amplitude collective oscillations. On the lower left panel,
we provide a zoomed up subplot of the  
survival probabilities in the range
$r = 1.$ to $1.1$. 
The neutrino evolution is similar to the bipolar neutrino oscillations
observed during usual collective neutrino oscillations. 
The only difference is that 
the oscillations are now occurring on scales determined by $\mu$ instead 
of $\omega$. It is also intersting to note that
the amplitude of collective oscillations decreases
  for larger matter density.

Fig. \ref{fig:alpha2} presents the same information as in Fig. \ref{fig:alpha5}
but for $\alpha=2$. For the matter density, we take 
$\lambda = 0$  \footnote{Note that for $\alpha<1$, the presence of matter is 
necessary to observe fast mode instabilities \cite{Chakraborty:2016lct, Abbar:2017pkh}.},
$0.3\mu$ and $\mu$. In contrast to the case of 
$\alpha<1$, for $\alpha>1$ there always seems to exist a sort of
stabilization with a relatively small fluctuation amplitude.
Despite this fact, it does not seem that any sort of flavor equilibrium is
generally reached since neither $P_{\nu_e \nu_e}=1/2$ nor 
$P_{\bar{\nu}_e \bar{\nu}_e}=1/2$ (except for the lowest left panel in which
$P_{\nu_e \nu_e} \approx 0.52$).
Note that the oscillation amplitude gets very tiny
for the maximum matter density.

In spite of the ongoing speculation,
fast modes do not necessarily lead to flavor equilibrium.
This speculation stems from the assumption that 
large amplitude fast flavor conversions could occur on scales 
(determined by $\mu$) much
smaller than the scales of collective modes (which were assumed
to be determined by $\omega$).
However, as it has been analytically shown in Ref. \cite{Abbar:2017pkh},
in the presence of fast modes in  neutrino gas,
  $\mu$ (or even $\lambda$) could
 play the role of $\omega$. Thus, the 
 collective scales are set by a combination of  $\mu$ and $\omega$ 
 (rather than only $\omega$) in the
 presence of fast modes. This means that the logic of 
 comparing scales mentioned above could totally fail.
 Then one might be tempted to  expect that the nature of the
evolution of fast modes (in the nonlinear regime) may not be 
much different from the one of the usual collective slow modes. 
  
Since the values of $\mu$ for which fast conversion modes occur
are very large, one might have to use a relatively large number of angle bins
 (at least
several thousands in our case) to reach convergence in the simulations.
 This simply
comes from the fact that for such large values of $\mu$,
 the neighbouring neutrino beams could experience 
quite different potentials during their evolution. This implies that the 
 angular distribution of the neutrino quantities could be completely 
 uneven. In Fig. \ref{fig:angular}, we present
the angular distributions of  the neutrino and antineutrino
survival probabilities  at different radii. 
Obviously, one needs a large enough number of angle bins to capture
all of the patterns in the angular distributions. 

\begin{figure} [t]
\begin{center}
\includegraphics*[width=.5\textwidth,trim=15 05 25 35,clip]{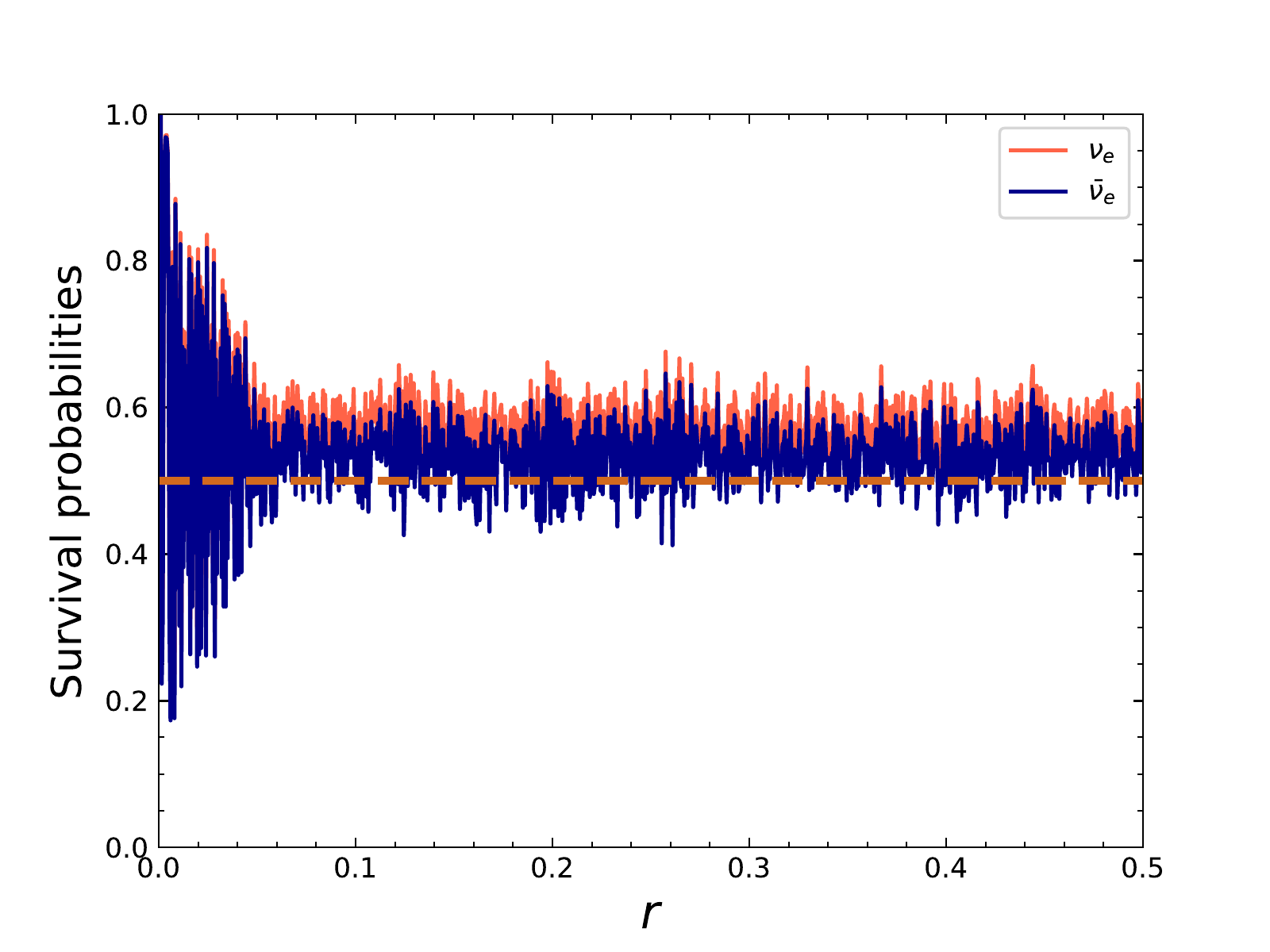}
\end{center}
\caption{ Artificial flavor equilibrium due to  insufficient number of 
angle bins. The angle-averaged  survival
probabilities of neutrinos (blue line) and antineutrinos (red line)
are plotted
as a function of the unitless distance $r=\mu z/10^4$
 for $\alpha=0.5$, $\mu= 10^5$, $\lambda = \mu$ and using $100$ angle bins.
To guide the eye, we have plotted the value $0.5$ (brown dashed  
line) which corresponds
to flavor equilibrium.} 
\label{fig:100ang}
\end{figure}

Although  flavor equilibrium could be hard to reach
in a valid treatment of the problem accounting well 
for numerical convergence,
a too small
number of angle bins may lead to an approximate  
artificial flavor equilibrium. This is shown explicitly in 
 Fig. \ref{fig:100ang} where only $100$ angle bins are used
 for $\alpha=0.5$,
 $\mu=10^5$ and $\lambda=\mu$ (to be compared
 with the lower right panel in Fig. \ref{fig:alpha5}). 

\begin{figure} [t]
\begin{center}
\includegraphics*[width=.5\textwidth,trim=20 05 25 35,clip]{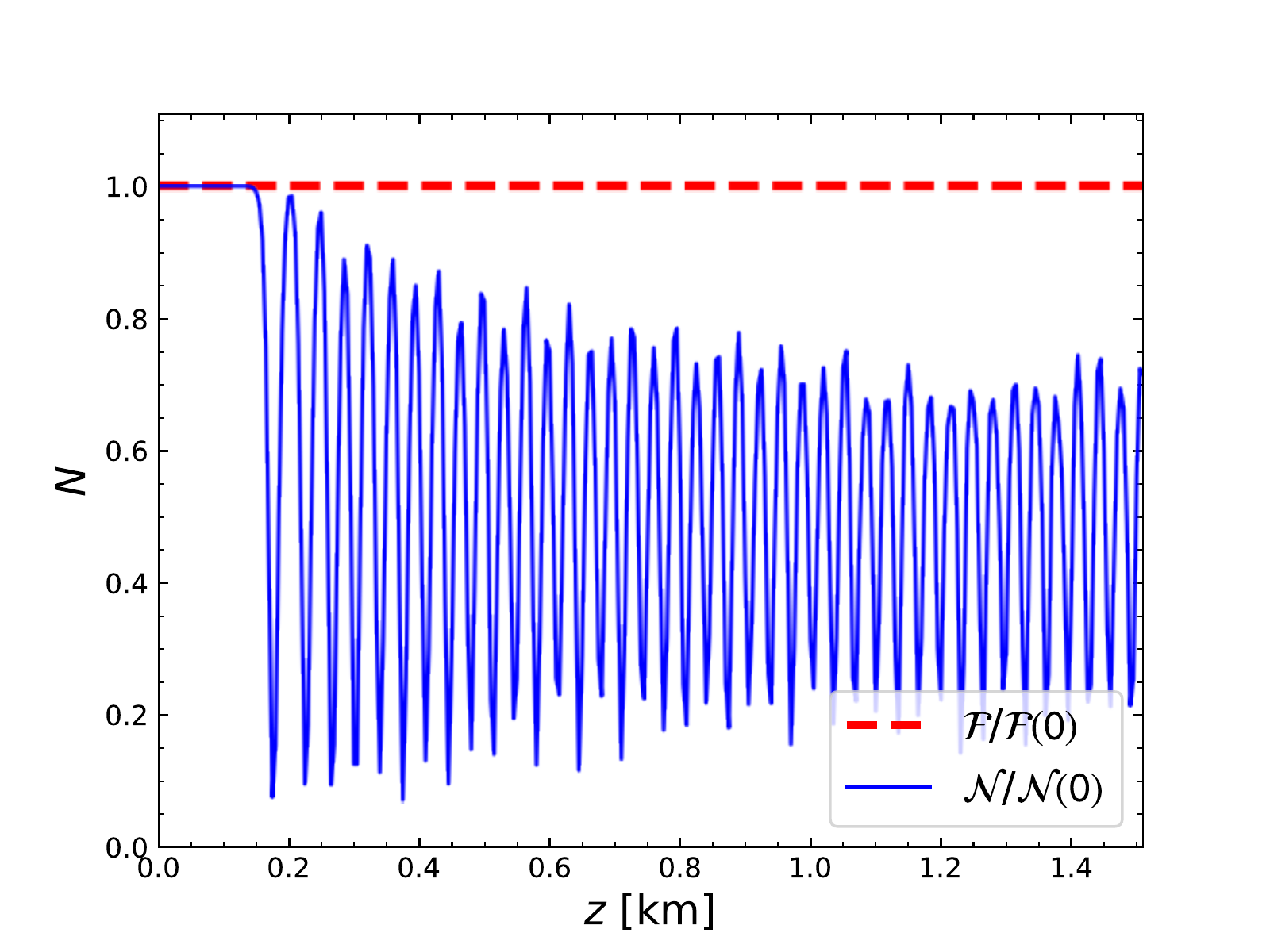}
\end{center}
\caption{The evolution of the net lepton number  $\mathcal{N}$ (blue
solid line) and the lepton number flux $\mathcal{F}$ (red dashed line).
The values are rescaled by the initial ones at $z=0$.} 
\label{fig:leptonN}
\end{figure}

Another remark concerns the neutrino lepton number 
conservation law  (for $\theta_{\mathrm{v}} \approx 0$)
within the comoving frame
\cite{Duan:2005cp, Duan:2007mv, Hannestad:2006nj}.
 Although if one considers only the temporal evolution of the
neutrino gas, the net lepton number
\begin{equation}
\begin{split}
\mathcal{N} = n_{\nu_e} \int &\big[ f_{\nu_e}(\vartheta') (\rho_{\nu_e\nu_e,\vartheta'}- \rho_{\nu_x\nu_x,\vartheta'})\\
  &- \alpha f_{\bar{\nu}_e}(\vartheta')
(\rho_{\bar{\nu}_e\bar{\nu}_e,\vartheta'} - \rho_{\bar{\nu}_x\bar{\nu}_x,\vartheta'} )\big]
   \    \mathrm{d}\vartheta' ,
\end{split}
 \end{equation}
 is conserved, it is
the flux of the  lepton number 
\begin{equation}
\begin{split}
\mathcal{F} = n_{\nu_e} \int &\big[ f_{\nu_e}(\vartheta') (\rho_{\nu_e\nu_e,\vartheta'} - \rho_{\nu_x\nu_x,\vartheta'})\\
  &- \alpha f_{\bar{\nu}_e}(\vartheta')
(\rho_{\bar{\nu}_e\bar{\nu}_e,\vartheta'} - \rho_{\bar{\nu}_x\bar{\nu}_x,\vartheta'})\big]
   \ \cos\vartheta'   \mathrm{d}\vartheta' ,
\end{split}
 \end{equation}
which is conserved if only the spatial evolution of
the neutrino gas is considred. One should note that 
the net lepton number is not conserved any more.
In particular, a nontrivial
evolution of the angular distribution of the $\nu$ELN could allow
for large variations of $\mathcal{N} $. We observed that this variations could be
larger for smaller $\nu_e$-$\bar{\nu}_e$ asymmetries. 
This is explicitly shown in Fig. \ref{fig:leptonN} 
where (rescaled) $\mathcal{F}$ and (rescaled) $\mathcal{N}$ are plotted
as a function of $z$ for $\alpha=0.8$ and $\lambda=0.3\mu$.  
Although (rescaled) $\mathcal{F}$ is expectedly conserved, 
the (rescaled) net lepton 
number $\mathcal{N}$ could vary by more than $50\%$. 
Though the extent to which the net lepton number
could be violated depends on the details of the physical quantities, 
it should be noted that one might not be, in principle, 
seriously limited by the conservation of the neutrino net lepton number.

In this study, we have assumed that  neutrino beams 
are only  propagating in the forward direction. 
However, if the backward propagating 
modes exist, 
one has to consider both spatial and temporal evolution
of  neutrinos   \cite{Capozzi:2017gqd}. 
Here we do not want to consider a multi-dimensional problem
which could be very tough in multiangle configuration.
However, to have an idea of
the temporal evolution of fast modes in the nonlinear regime,
we considered a homogenous non-stationary  neutrino gas so that the time
is enough (and the relevant parameter) to describe the evolution of the 
system. The neutrinos were assumed to be emitted in
a multiangle scenario with emission angles $\vartheta$ 
within the range $[-\vartheta_{\rm{max}}, \vartheta_{\rm{max}}]$.
One then has to solve an EOM which is similar to  Eq. (\ref{eq:EOM})
except that $\cos\vartheta \  \partial_z$ must be 
replaced by $\partial_t$. 
We repeated our simulations for the temporal evolution of fast modes  
 and we did not observe any qualitative difference for $\alpha>1$
for which we observed fast modes in the neutrino gas (right panel of Fig. \ref{fig:time}). 
For the $\nu_e$ dominated case, we did not observe any temporal
instability for the SN-motivated parameters. However,  fast modes
could exist if one exchanges the initial angular distributions of $\nu_e$ and $\bar{\nu}_e$
with each other
so that the opening angle of $\nu_e$ becomes smaller than that of  $\bar{\nu}_e$.
Although this is not currently thought 
to be realistic in the SN environment, we still think it is useful
to be discussed (left panel of Fig. \ref{fig:time}) since it  helps convey the message that the 
 logic of comparing scales could fail.

By studying a one dimensional spatial/temporal schematic model,
we have shown that fast modes do
not necessarily lead to flavor equilibrium
or even flavor stabilization in the \emph{two-flavor} scenario. 
The non-occurrence of flavor equilibrium makes the problem of flavor evolution in
dense neutrino media more involved.
Obviously, it remains to be investigated if multi-dimensional
models can induce flavor equilibrium and a degeneracy of the neutrino spectra which
would remarkably simplify the issue of  neutrino evolution in core-collapse supernovae or accretion
disks around compact objects such as black holes or binary neutron star merger remnants.
In conclusion, unless future studies demonstrate that such an equilibration can occur, 
the present work shows that the study of the evolution of a dense neutrino gas requires a seven
dimensional non-linear description, implementing symmetry breaking and possible small
scale instabilities due to fast modes.

\section*{Acknowledgments}
We would like to thank Huaiyu Duan for useful discussions. 
We also acknowledge support from ”Gravitation et physique 
fondamentale” (GPHYS) of the Observatoire de Paris.

\section*{References}
\bibliographystyle{elsarticle-num}
\bibliography{fast}

\end{document}